\documentclass[]{sig-alternate-10pt}

\usepackage{color}
\usepackage{amsmath}
\usepackage{subfig}
\usepackage{graphicx}
\usepackage{hhline}
\usepackage{textgreek}

\begin{document}

\title{On the anonymizability of mobile traffic datasets}

\numberofauthors{2} 
%
\author{
%
%
\alignauthor
	Marco Gramaglia\\
	\affaddr{CNR - IEIIT}\\
	\affaddr{Corso Duca degli Abruzzi 24}\\
	\affaddr{10129 Torino, Italy}\\
	\email{marco.gramaglia@ieiit.cnr.it}
\alignauthor
	Marco Fiore\\
	\affaddr{CNR - IEIIT}\\
	\affaddr{Corso Duca degli Abruzzi 24}\\
	\affaddr{10129 Torino, Italy}\\
	\email{marco.fiore@ieiit.cnr.it}
}

\maketitle

\begin{abstract}
Preserving user privacy is paramount when it comes to
publicly disclosed datasets that contain fine-grained
data about large populations.
The problem is especially critical in the case of mobile
traffic datasets collected by cellular operators, as they
feature elevate subscriber trajectory uniqueness	
and
they are resistant to anonymization through spatiotemporal
generalization.
In this work, we investigate the
$k$-anonymizability of
trajectories
in two large-scale mobile traffic datasets,
by means of a novel dedicated measure.
Our results are in agreement with those of previous
analyses, however they also provide additional insights
on the reasons behind the poor anonimizability of mobile
traffic datasets.
As such, our study is a step forward in the direction of
a more robust dataset anonymization.
\end{abstract}

\section{Introduction}
\label{sec:intro}


Public disclosure of datasets containing {\it micro-data}, i.e.,
information on precise individuals, is an increasingly frequent practice.
Such datasets are collected in a number of different ways, including surveys,
transaction recorders, positioning data loggers, mobile applications, and
communicaiton network probes. They yield fine-grained data about large
populations that has proven critical to seminal studies in a number of
research fields.

However, preserving user privacy in publicly accessible micro-data datasets
is currently an open problem.
Publishing an incorrectly anonymized dataset may disclose sensible information
about specific users. This has been repeatedly proven in the past. One of the
first and best known attempts at re-identification of badly anonymized datasets
was carried out by then MIT graduate student Latanya Sweeney~\cite{sweeney02kanon,ohm2009broken}
in 1996. By using a database of medical records
released by an insurance company and the voter roll for the city of Cambridge (MA),
purchased for 20 US dollars, Dr. Sweeney could successfully re-identify the full
medical history of the then governor of Massachusetts, William Weld. She even
sent the governor full health records, including diagnoses and prescriptions,
to his office.
A later, yet equally famous experiment was performed by
Narayanan \emph{et al.}~\cite{narayanan08robust}
on a dataset released by Netflix for a data-mining contest, which was
cross-correlated with a web scraping of the popular IMDB website.
The authors were able to match two users from both datasets revealing, e.g.,
their political views.

Mobile traffic datasets include micro-data collected at different locations
of the cellular network infrastructure, concerning the movements and traffic
generated by thousands to millions of subscribers, typically for long timespans
in the order of weeks or months.
They have become a paramount instrument in large-scale analyses across disciplines
such as sociology, demography, epidemiology, or computer science.
Unfortunately, mobile traffic datasets may also be prone to attacks on individual
privacy. Specifically, they suffer from the following two issues.

\begin{enumerate}
\item {\bf Elevate uniqueness.} Mobile subscribers have very distinctive
patterns that often make them unique even within a very large population.
Zang and Bolot~\cite{zang11large} showed that 50\% of the mobile subscribers
in a 25 million-strong dataset could be uniquely detected with minimal
knowledge about their movement patterns, namely the three locations they
visit the most frequently.
The result was corroborated by de Montjoye \emph{et al.}~\cite{deMontjoye13unique},
who demonstrated how an individual can be pinpointed among 1.5 million other
mobile customers with a probability almost equal to one, by just knowing five
random spatiotemporal points contained in his mobile traffic data.
\end{enumerate}

Uniqueness does not implies identifiability, since the sole knowledge of
a unique subscriber trajectory cannot disclose the subscriber's identity.
Building that correspondence requires instead sensible side information
and cross-database analyses similar to those carried out on medical or
Netflix records.
To date, there has been no actual demonstration of subscriber re-identification
from mobile traffic datasets using such techniques -- and our study does not
change that situation.
Still, uniqueness may be a first step towards re-identification, and whether
this represents a threat to user privacy is an open topic for
discussion~\cite{cavoukian14a,narayanan14}.

In such a context, the standard, safe approach to ensure data confidentiality
relies on non-technical solutions, i.e., non-disclosure agreements that well
define the scope of the activities (e.g., fundamental research only) carried
out on the datasets, and that prevent open disclosure of the data or results
without prior verification by the relevant authorities. This is, for instance,
the solution adopted in the case of the mobile traffic information we will
consider in Sec.\,\ref{sec:datasets}.

Clearly, this practice can strongly limit the availability of mobile traffic
datasets, as well as the reproducibility of related research.
Mitigating the uniqueness of subscriber trajectories becomes then a very
desirable facility that can entail more privacy-preserving datasets, and
favor their open circulation.
It is however at this point that the second problem of mobile traffic datasets
comes into play.

\begin{enumerate}
\item[2.] {\bf Low anonymizability.} 
The legacy solution to reduce uniqueness in micro-data datasets is generalization
and suppression.
However, previous studies showed that
blurring users in the crowd, by reducing the spatial and temporal granularity
of the data, is hardly a solution in the case of mobile traffic datasets.
Zang and Bolot~\cite{zang11large} found that reliable anonymization is attained
only under very coarse spatial aggregation, namely when the mobile subscriber
location granularity is reduced to the city level.
Similarly, de Montjoye \emph{et al.}~\cite{deMontjoye13unique} proved that a
power-law relationship exists between uniqueness
and spatiotemporal
aggregation of mobile traffic. This implies that privacy is increasingly hard
to ensure as the resolution of a dataset is reduced.
In conclusion, not only mobile traffic datasets 
yield highly unique trjectories,
but the latter are also hard to anonymize.
Ensuring individual privacy risks to
lower the level of detail of such datasets to the point that they are not informative
anymore.
\end{enumerate}

In this work, we aim at better investigating the reasons behind such inconvenient
properties of mobile traffic datasets. We focus on anonymizability, since it is
a more revealing feature: multiple datasets that
feature similar trajectory
uniqueness
may be more or less difficult to anonymize.
Attaining our objective brings along the following contributions:
(i) we define a measure of the level of anonymizability of mobile traffic datasets,
in Sec.\,\ref{sec:metric};
(ii) we provide a first assessment of the anonymizability of two large-scale
mobile traffic datasets, in Sec.\,\ref{sec:datasets};
(iii) we unveil the cause of elevate uniqueness
and poor anonymizability
in such datasets, i.e., the heavy tail of the temporal diversity among subscriber
mobility patterns, in Sec.\,\ref{sec:results}.
Finally, Sec.\,\ref{sec:conc} concludes the paper.



\section{How anonymizable is your\\mobile traffic fingerprint?}
\label{sec:metric}

In this section, we first define in a formal way the problem of user
uniqueness in mobile traffic datasets, in Sec.\,\ref{sub:problem}.
Then, we introduce the proposed measure of anonymizability, in Sec.\,\ref{sub:measure}.

\subsection{Our problem}
\label{sub:problem}

In order to properly define the problem we target, we need to
introduce the notion of mobile traffic fingerprint that is at
the base of the mobile traffic dataset format. We also need to
specify the type of anonymity we consider -- in our case,
$k$-anonymity. Next, we discuss these aspects of the problem.

\begin{table}[tb]
\small
\centering
\caption{Standard micro-data database format.}
\vspace*{-3pt}
\label{tab:db_std} 
\renewcommand{\arraystretch}{1.1}
\setlength{\tabcolsep}{4pt}
\begin{tabular}{|l|l|l|l|l|l|l}
\hline
Pseudo-id & Gender & Age & ZIP & Degree & Income & \dots \\
\hline
\hline
00013701 & Male & 21 & 77005 & Bachelor & 13,000 & \dots \\
08936402 & Male & 37 & 77065 & Master's & 90,000 & \dots \\
42330327 & Female & 60 & 89123 & High School &  46,000 & \dots \\
\dots & \dots & \dots & \dots & \dots & \dots & \dots \\
\end{tabular}
\end{table}

\begin{table}[tb]
\small
\centering
\caption{Mobile traffic database format.}
\vspace*{-3pt}
\label{tab:db_mt} 
\renewcommand{\arraystretch}{1.1}
\setlength{\tabcolsep}{4pt}
\begin{tabular}{|c|l|l|l|l|l|l|l|}
\hline
Pseudo-id & \multicolumn{7}{|l|}{Spatiotemporal samples (fingerprint)} \\
\hhline{--------}
a & $c_1$,8 & $c_2$,14 & $c_3$,17 & \multicolumn{4}{l}{} \\
\hhline{--------}
b & $c_4$,8 & $c_5$,15 & $c_6$,15 & \dots & $c_{13}$,15 & $c_{14}$,16 & $c_{15}$,17 \\
\hhline{--------}
c & $c_{16}$,7 & $c_{17}$,20 & \multicolumn{5}{l}{} \\
\hhline{---~~~~~}
\dots & \dots & \multicolumn{6}{l}{} \\
\hhline{--~~~~~~}
\end{tabular}
\end{table}

\subsubsection{Mobile traffic fingerprint and dataset}

Traditional micro-data databases are structured into matrices where each
row maps to one individual, and each column to an {\it attribute}. An example
is provided in Tab.\ref{tab:db_std}.
Individuals are associated to one {\it identifier}, i.e., a value that
uniquely pinpoints the user across datasets (e.g., his complete name,
social number, or passport number). Since identifiers allow direct
identification and immediate cross-database correlation, they are never
disclosed.
Instead, they are replaced by a {\it pseudo-identifier}, which is again
unique for each individual, but changes across datasets (e.g., a random
string substituting the actual identifier).
Then, standard re-identification attacks leverage {\it quasi-identifiers},
i.e., a sequence of known attributes of one user (e.g., the age, gender,
ZIP code, etc.) to recognize the user in the dataset. If successful, the
attacker has then access to the complete record of the target user.
This knowledge can directly include sensitive attributes, i.e., items
that should not be disclosed because they may pertain to the personal
sphere of the individual (e.g., diseases, political or religious views,
sexual orientation, etc.).
It can also be exploited for further cross-database correlation so as
to extract additional private information about the user.

\begin{figure*}[tb]
\centering
\subfloat[Initial]{\label{fig:map_def}
	\includegraphics[width=0.31\textwidth]{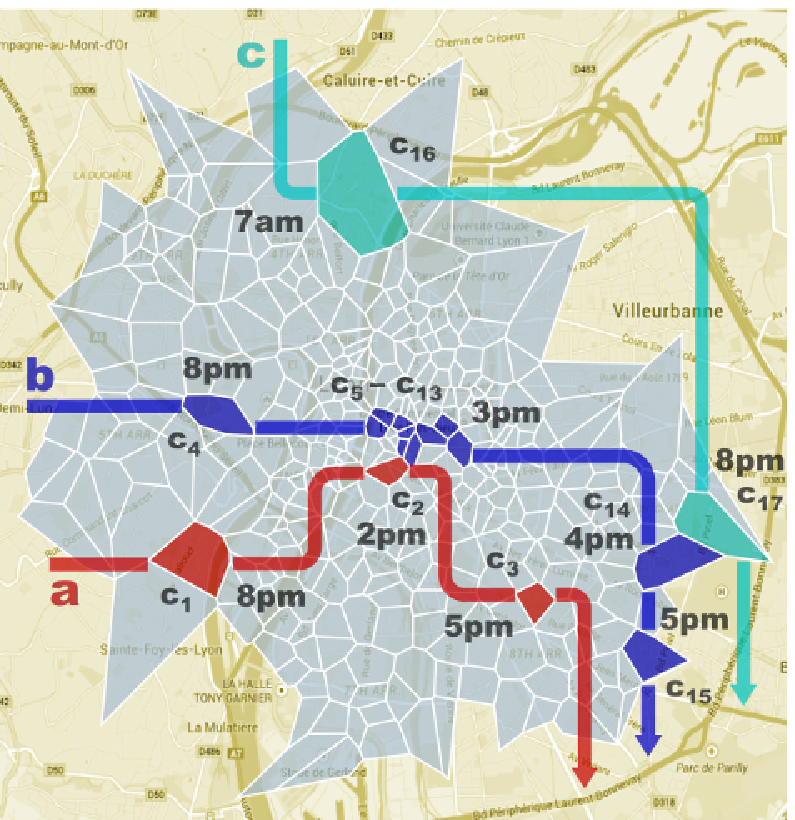}
}
\hspace*{3pt}
\subfloat[Aggregated]{\label{fig:map_arr-2h}
	\includegraphics[width=0.31\textwidth]{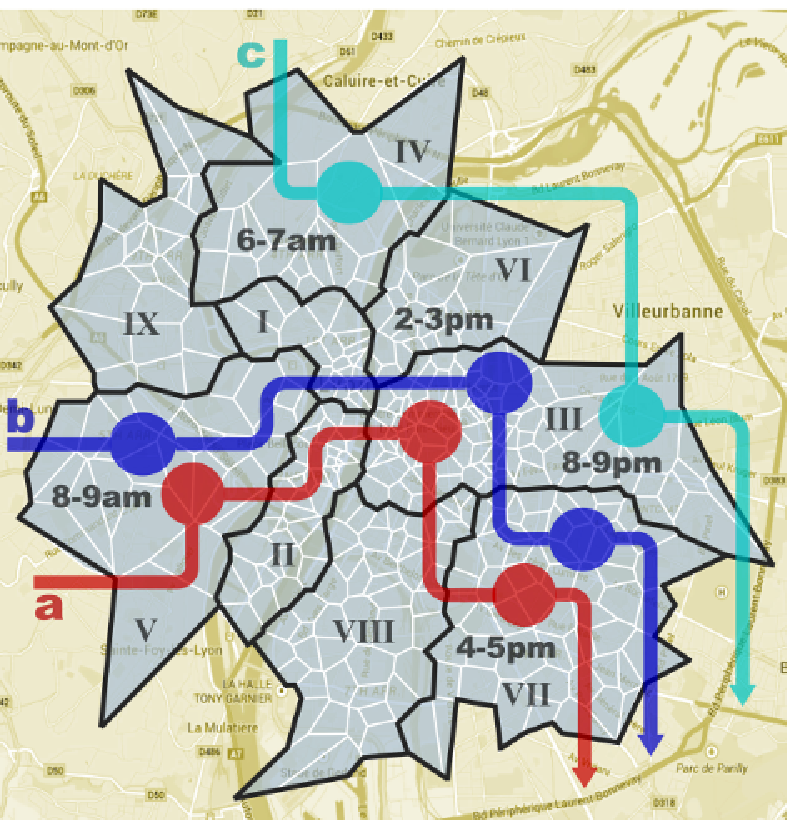}
}
\hspace*{3pt}
\subfloat[More aggregated]{\label{fig:map_half-12h}
	\includegraphics[width=0.31\textwidth]{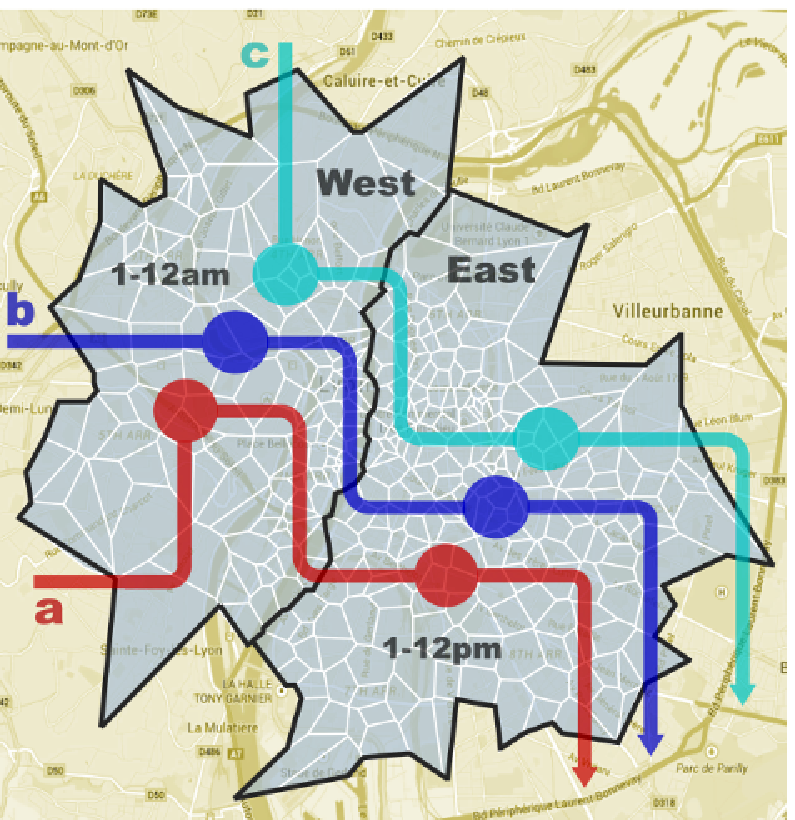}
}
\caption{Example of mobile traffic fingerprints of three subscribers.
(a) Initial dataset granularity: user locations are represented at cell
level, and the temporal information has a hourly precision.
(b) First aggregation level: positions are recorded at each neighborhood,
and the time granularity is reduced to two hours.
(c) Second aggregation level: location data is limited to Eastern or
Western half of the city, and the time information is merged over 12 hours.}
\label{fig:map}
\end{figure*}

The same model directly applies to the case of mobile traffic datasets.
However, the database semantics make all the difference here: while mobile
users are the obvious individuals whose privacy we want to protect, attributes
are now sequences of spatiotemporal samples. Each sample is the result of an
event that the cellular network associated to the user.
An illustration is provided in Fig.\,\ref{fig:map_def}, which portrays the
trajectories of three mobile customers, denoted with pseudo-identifiers
$a$, $b$, and $c$, respectively, across an urban area.
User $a$ interacts with the radio access infrastructure
at 8 am, while he is in cell $c_1$ along his trajectory.
Then, he triggers additional mobile traffic activities at 2 pm, while located
in a cell $c_2$ in the city center, and at 5 pm, from a cell $c_3$ in
the South-East city outskirts. The same goes for users $b$ and $c$.
All these spatiotemporal samples are recorded by the mobile operator%
\footnote{The actual precision of the information recorded, both in space
and in time, can depend significantly on the nature of the probes used
by the operator. Typically, probes located closer to the radio access can
capture more events at a finer granularity, but require more extensive
deployments to attain a similar coverage than lower-precision probes
located in the mobile network core. In all cases, our discussion is
independent of the mobile traffic data collection technique, and all
the analyses performed in this work can be applied to any type of
mobile traffic data.}
and constitute the {\it mobile traffic fingerprint} of the user.
The resulting database has a format such as that in Tab.\ref{tab:db_mt},
where subscriber identifiers are replaced by pseudo-identifiers, and
each element of a user's fingerprint is a cell and hourly timestamp pair.

\subsubsection{$k$-anonymity in mobile traffic}

In order to preserve user privacy in micro-data, one has to ensure
that no individual can be uniquely pinpointed in a dataset.
This principle has led to the definition of multiple notions of non-uniqueness,
such as $k$-anonymity~\cite{sweeney02kanon},
$l$-diversity~\cite{Machanavajjhala07ldivers}
and $t$-closeness~\cite{li07tclose}.
Among those, $k$-anonymity is the baseline criterion, to which
$l$-diversity or $t$-closeness add further security layers that cope with
sensitive attributes or cross-database correlation.
More precisely, $k$-anonymity ensures that, for each individual, the set
of attributes (or its quasi-identifier subset) is identical to that of
at least other $k$-1 users. In other words, each individual is always
hidden in a crowd of $k$, and thus he cannot be uniquely identified among
such other users.

Granting $k$-anonymity in micro-data databases implies generalizing and
suppressing data. As an example, in order to ensure $2$-anonymity on the
age and ZIP code attributes for the first user in Tab.\,\ref{tab:db_std} ,
one can aggregate the age in twenty-year ranges, and the ZIP codes in
three-number ranges: both the first and second user end up with a
\texttt{(20,40)} age and \texttt{770**} ZIP code, which makes them
both $2$-anonymous.
Clearly, the process is lossy, since the information granularity is
reduced.
Many efficient algorithms have been proposed that achieve $k$-anonymity
in legacy micro-data databases, while minimizing information loss~\cite{ayala14kanon}.

Also in mobile traffic datasets, $k$-anonymity is regarded as a best
practice, and data aggregation is the common approach to achieve
it~\cite{zang11large,deMontjoye13unique}.
In this case, one has to ensure that the fingerprint of each subscriber
is identical to that of at least other $k$-1 mobile users in the same
dataset.
We remark that previous works have typically considered a model of
attacker who only has partial knowledge of the subscribers' fingerprints,
e.g., most popular locations~\cite{zang11large} or random samples~\cite{deMontjoye13unique}.
In order to counter such a attack model, a partial $k$-anonymization,
targeting the limited information owned be the attacker, would be
sufficient.
However, we are interested in a general solution, so we do not make
any assumption on the precise knowledge of the attacker, which can
be diverse and possibly broad.
Thus, $k$-anonymizing the whole fingerprint of each subscriber in
the mobile traffic dataset is the only way to deterministically
ensure mobile user privacy.

Both spatial and temporal aggregations can be leveraged to attain
this goal.
Examples are provided in Fig.\,\ref{fig:map_arr-2h} and Fig.\,\ref{fig:map_half-12h}.
In Fig.\,\ref{fig:map_arr-2h}, cells are aggregated in large sets that roughly
map to the nine major neighborhoods of the urban area; also, time is
aggregated in two-hour intervals.
The reduction of spatiotemporal granularity allows $2$-anonymizing
mobile users $a$ and $b$: both have now a fingerprint composed by
samples \texttt{(V,8-9)}, \texttt{(III,14-15)}, and \texttt{(VII,16-17)}.
User $c$ has instead a different footprint, with samples
\texttt{(IV,6-7)} and \texttt{(III,20-21)}.
If we need to $3$-anonymize all three mobile customers in the example,
then a further generalization is required, as in Fig.\,\ref{fig:map_half-12h}.
There, the metropolitan region is divided in West and East halves, and
only two time intervals, before and after noon, are considered.
The result is that all subscribers $a$, $b$, and $c$ have identical
fingerprints \texttt{(West,1-12)} and \texttt{(East,13-24)}.
Clearly, this level of anonymization comes at a high cost in terms
of information loss, as the location data is very coarse both in
space and time.

This is precisely the problem of low anonymizability of mobile traffic
datasets unveiled by previous works~\cite{zang11large,deMontjoye13unique}:
even guaranteeing $2$-anonymization in a very large population requires
severe reductions of the spatiotemporal granularity, which limits the
usability of the data.

\subsection{A measure of anonymizability}
\label{sub:measure}

We intend to devise a measure of anonymizability that is based on
the $k$-anonymity criterion. Thus, our proposed measure evaluates the
effort, in terms of data aggregation, needed to make a user
indistinguishable from $k$-1 other subscribers.

We start by defining the distance between two spatiotemporal samples
in the mobile traffic fingerprints of two mobile users.
Each sample is composed of a spatial information (e.g., the cell
location) and a temporal information (e.g., the timestamp).
The distance must keep into account both dimensions.
A generic formulation of the distance between the $i$-th sample
of $a$'s fingerprint, $(s^a_i,t^a_i)$, and the $j$-th sample of
$b$'s fingerprint, $(s^b_j,t^b_j)$, is
\begin{equation}
d_{ab}(i,j) = w_s \delta_s\left(s^a_i,s^b_j\right) + w_t \delta_t\left(t^a_i,t^b_j\right).
\label{eqn:sampledist}
\end{equation}
Here, $\delta_s$ and $\delta_t$ are functions that determine the
distance along the spatial and temporal dimensions, respectively.
The former thus operates on the spatial information in the two
samples, $s^a_i$ and $s^b_j$, and the latter on the temporal
information, $t^a_i$ and $t^b_j$.
The factors $w_s$ and $w_t$ weigth the spatial and temporal
contributions in (\ref{eqn:sampledist}). In the following, we
will assume that the two dimensional have the same importance,
thus $w_s = w_t = 1/2$.

We shape the $\delta_s$ and $\delta_t$ functions by considering
that both spatial and temporal aggregations induce a loss of
information that is linear with the decrease of granularity.
However, above a given spatial or temporal threshold, the information
loss is so severe that the data is not usable anymore.
As a result, the functions can be expressed as
\begin{align}
\delta_s\left(s^a_i,s^b_j\right) & =
\begin{cases}
	\displaystyle
	\frac{dist\left(s^a_i,s^b_j\right)}{\delta^{max}_s} & \text{if } dist\left(s^a_i,s^b_j\right)\leq \delta^{max}_s \\
	1 & \text{otherwise},
\end{cases}
\label{eqn:deltas}
\end{align}
and
\begin{align}
\delta_t\left(t^a_i,t^b_j\right)  & =
\begin{cases}
	\displaystyle
	\frac{|t^a_i-t^b_j|}{\delta^{max}_t} & \text{if } |t^a_i-t^b_j|\leq \delta^{max}_t \\
	1 & \text{otherwise}.
 \end{cases}
 \label{eqn:deltat}
\end{align}
In (\ref{eqn:deltas}), $dist\left(s^a_i,s^b_j\right) = |s^a_i.x-s^b_j.x| + |s^a_i.y-s^b_j.y|$
is the \emph{Taxicab distance}~\cite{krause1973taxicab} between the
spatial components of the samples, whose coordinates are denoted as $x$ and
$y$ in a valid map projection system.
Both functions fulfill the properties of distances, i.e., are positive
definite, symmetric, and satisfy the triangle inequality.
They range from $0$ (samples are identical from a spatial or temporal
viewpoint) to $1$ (samples are at or beyond the maximum meaningful
aggregation threshold).
Concerning the values of the thresholds, in the following we will consider
that the aggregation limits beyond which the information deprivation is
excessive are 20~km for the spatial dimension (i.e., the size of a city,
beyond which all intra-urban movements are lost) and 8~hours (beyond which
the night, working hours, and evening periods are merged together).

The sample distance in (\ref{eqn:sampledist}) can be used to define the
distance among the whole fingerprints of two mobile subscribers $a$ and $b$,
as
\begin{align}
\Delta_{ab}  & =
\begin{cases}
	\displaystyle
	\frac{1}{n_a}\sum_{h=1}^{n_a}\min_{k=1,\dots,n_b} d_{ab}(h,k) & \text{if } n_a \geq n_b \\
	\displaystyle
	\frac{1}{n_b}\sum_{h=1}^{n_b}\min_{k=1,\dots,n_a} d_{ab}(k,h) & \text{otherwise}.
 \end{cases}
 \label{eqn:fingerdist}
\end{align}
Here, $n_a$ and $n_b$ are the cardinalities of the fingerprints of $a$ and $b$,
respectively. The expression in (\ref{eqn:fingerdist}) takes the longer fingerprint
between the two, and finds, for each sample, the sample at minimum distance
in the shorter fingerprint. The resulting $\Delta_{ab}$ is the average among
all such sample distances, and $\Delta_{ab} = \Delta_{ba}$, $\forall a,b$.

The measure of anonymizability of a generic mobile user $a$ can be mapped,
under the $k$-anonymity criterion, to the average distance of his fingerprint
from those of the nearest $k$-1 other users. Formally
\begin{equation}
\Delta^k_a = \frac{1}{k-1} \sum_{b\in\mathbb{N}^{k-1}_a} \Delta_{ab},
\label{eqn:delta}
\end{equation}
where $\mathbb{N}^{k-1}_a$ is the set of $k-1$ users $b$ with the smallest
fingerprint distances to that of $a$.

The expression in (\ref{eqn:delta}) returns a measure $\Delta^k_a \in [0,1]$ that
indicates how hard it is to hide subscriber $a$ in a the crowd of $k$ users.
If $\Delta^k_a = 0$, then the user is already $k$-anonymized in the dataset.
If $\Delta^k_a = 1$, the user is completely isolated, i.e., no sample in the
fingerprints of all other subscribers is within the spatial and temporal thresholds,
$\delta^{max}_s$ and $\delta^{max}_t$, from any samples of $a$'s fingerprint.


\section{Two mobile traffic use cases}
\label{sec:datasets}

We employ the proposed measure to assess the level of anonymizability
of fingerprints present in two mobile traffic datasets released by
Orange in the framework of the Data for Development Challenge.
In order to allow for a fair comparison, we preprocessed the datasets
so as to make them more homogeneous.
\begin{itemize}
\item \textbf{Ivory Coast.}
Released for the 2012 Challenge, this dataset describes five months of
Call Detail Records (CDR) over the whole the African nation of Ivory Coast.
We used the high spatial resolution dataset, containing the complete
spatio temporal trajectories for a subset of 50,000 randomly selected users
that are changed every two weeks. Thus, the dataset contains information
about 10 2-weeks periods overall. We performed a preliminary screening,
discarding the most disperse trajectories, keeping the users that have at
least one spatio-temporal point per day. Then, we merged all the user that
met this criteria in a single dataset, so as to achieve a meaningful size
of around 82,000 users. This dataset is indicated as \texttt{d4d-civ} in the
following.
\item \textbf{Senegal.}
The 2014 Challenge dataset is derived from CDR collected over the whole
Senegal for one year.
We used the fine-grained mobility dataset, containing a randomly selected
subset of around 300,000 users over a rolling 2-week period, for a total of
25 periods.
We did not filter out subscribers, since the dataset is already limited to
users that are active for more than 75\% of the 2-week time span.
In our study, we consider one representative 2-week period among those available.
This dataset is referred to as \texttt{d4d-sen} in the following.
\end{itemize}
In both the mobile traffic datasets, the information about the user position%
\footnote{The spatial information maps to the antenna location in \texttt{d4d-civ},
and to a random point within the voronoi cell associated to the antenna in \texttt{d4d-sen}.}
is provided as a latitude and longitude pair. We projected the latter in a
two-dimensional coordinate system using the Lambert azimuthal equal-area
projection.
We then discretize the resulting positions on a 100-m regular grid, which
represents the maximum spatial granularity we consider%
\footnote{At 100-m spatial granularity, each square cell contains at most
one antenna or voronoi location from the original dataset. In other words,
this discretization does not implies any spatial aggregation.}.
As far as the temporal dimension is concerned, the maximum precision granted
by both datasets is one minute, and this is also our finest time granularity.

\section{Results}
\label{sec:results}

The measure of anoymizability in (\ref{eqn:delta}) can be intended as a dissimilarity
measure, and employed in legacy definitions used to understand micro-data
database sparsity, e.g., \texttt{(\textepsilon,\textdelta)-sparsity}~\cite{narayanan08robust}.
However, these definitions are less informative than the complete distribution of
the anonymizability measure. Thus, in this section, we employ Cumulative
Distribution Functions (CDF) of the measure in (\ref{eqn:delta}) in order to assess
the anonymizability of the two datasets presented before.

\subsection{The good: anonymity is close to reach}

\begin{figure}
\centering
	\includegraphics[width=0.8\columnwidth]{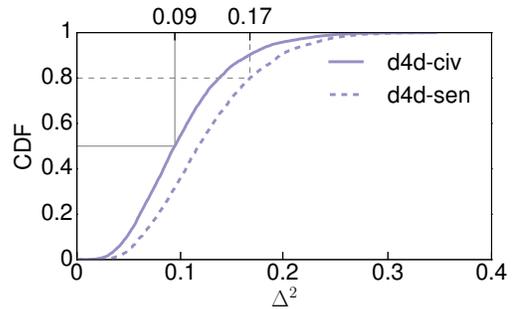}
\caption{CDF of the anonymizability measure, under the $2$-anonymity criterion,
in the \texttt{d4d-civ} and \texttt{d4d-sen} mobile traffic datasets.}
\label{fig:d4d_cdf}
\end{figure}

Our basic result is shown in Fig.\,\ref{fig:d4d_cdf}. The plot portrays the
CDF of the anonymizability measure computed on all users in the two reference
mobile traffic datasets, \texttt{d4d-civ} and \texttt{d4d-sen}, when considering
$2$-anonymity as the privacy criterion.

We observe that the two curves are quite similar, and both are at zero in the
x-axis origin. This means that no single mobile subscriber is $2$-anonymous in
either of the original datasets.
Since similar observations were made on different data~\cite{zang11large,deMontjoye13unique},
our results seem to confirm that the elevate uniqueness of subscriber trajectories
is an intrinsic property of any mobile traffic dataset, and not just a specificity
of those we analyse in this study.

More interestingly, the probability mass gathered in both cases in the $0.1$-$0.2$ range,
i.e., it is quite close to the origin. This is good news, since it implies
that the average aggregation effort needed to achieve $2$-anonymity is not
elevate.
As an example, 50\% of the users in the \texttt{d4d-civ} dataset have a measure
$0.09$ or less, which maps, on average, to a combined spatiotemporal aggregation
of less than one km and little more than 20 minutes. In other words, the result
seems to suggest that half of the individuals in the dataset can be $2$-anonymized
if the spatial granularity is decreased to 1~km, and the temporal precision is
reduced to around 20 minutes.
Similar considerations hold in the \texttt{d4d-sen} case, where, e.g., 80\%
of the dataset population has a measure $0.17$ or less. Such a measure is the
result of average spatial and temporal distances of 1.7~km and 41 minutes from
$2$-anonymity.

\begin{figure}
\centering
\hspace*{-5pt}
\subfloat[d4d-civ]{
	\includegraphics[width=0.52\columnwidth]{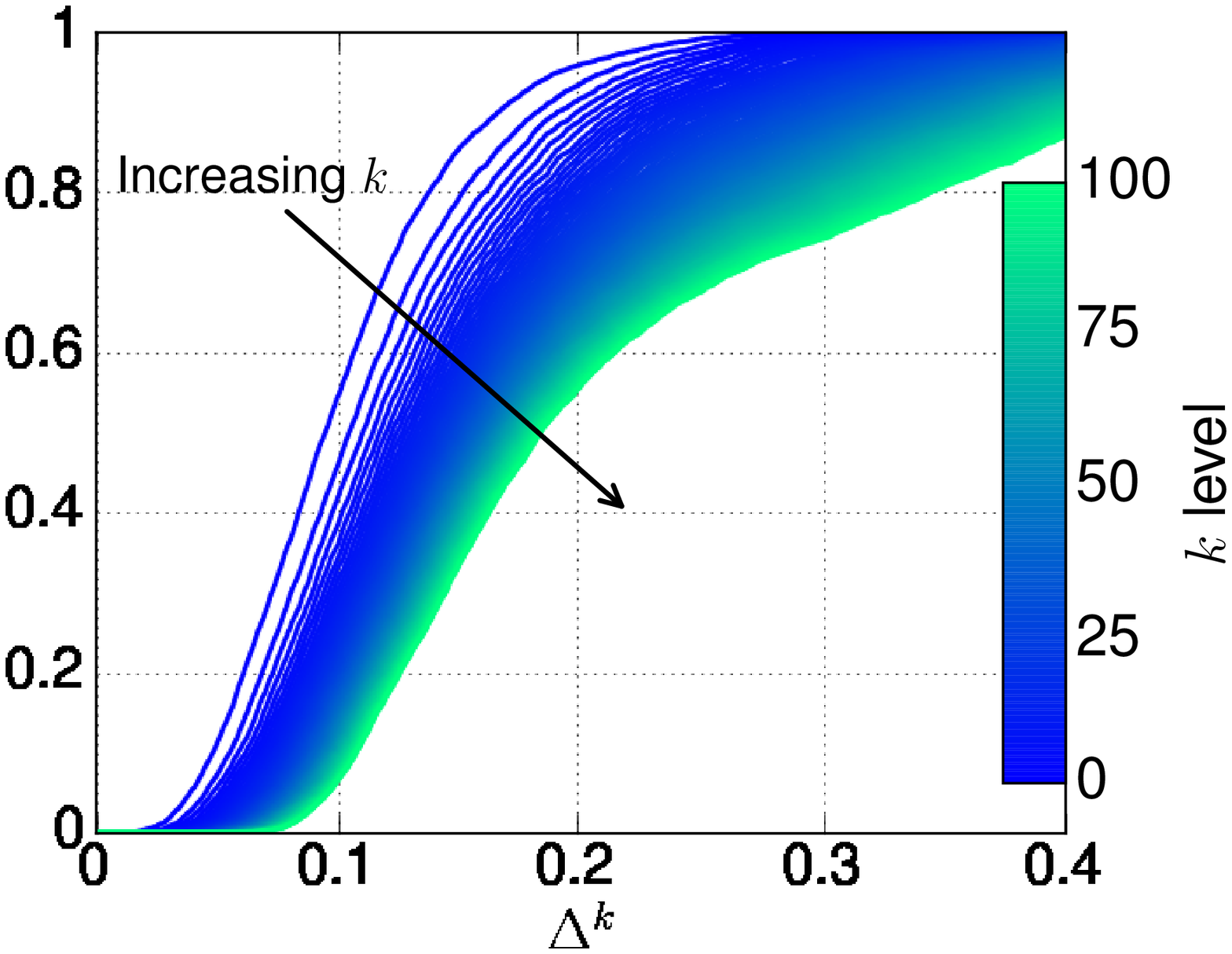}
	\label{fig:d4d_vark_civ}
}
\hspace*{-5pt}
\subfloat[d4d-sen]{
	\includegraphics[width=0.52\columnwidth]{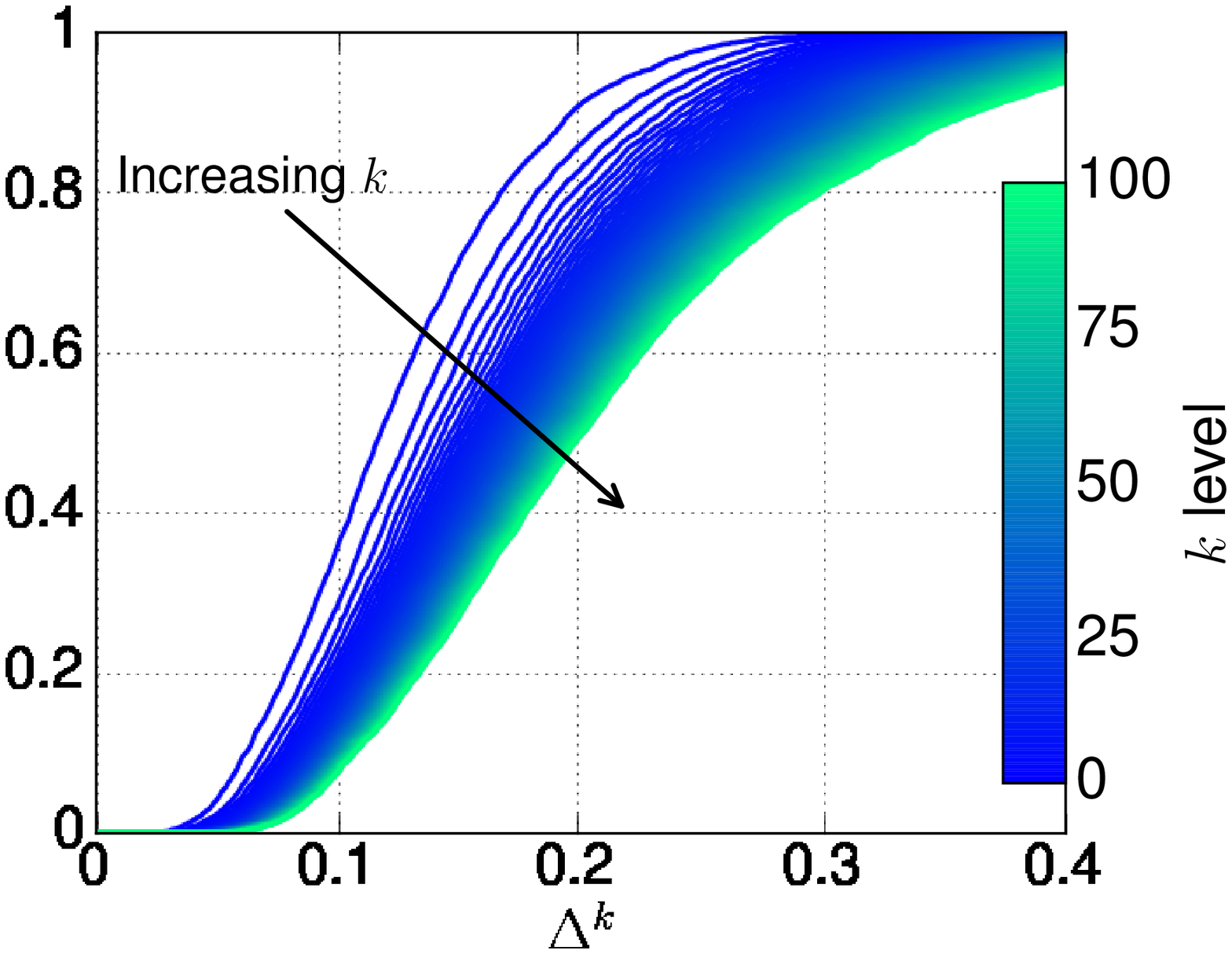}
	\label{fig:d4d_vark_sen}
}
\hspace*{-5pt}
\caption{CDF of the anonymizability measure, for varying $k$ of the
$k$-anonymity criterion, in the \texttt{d4d-civ} and \texttt{d4d-sen}
mobile traffic datasets.}
\label{fig:d4d_vark}
\end{figure}

One may wonder how more stringent privacy requirements affect these results.
Fig.\,\ref{fig:d4d_vark} shows the evolution of the anonymizability of
the two datasets when $k$ varies from $2$ to $100$.
As expected, higher values of $k$ require that a user is hidden in a larger
crowd, and thus shift the distributions towards the right, implying the
need for a more coarse aggregation.
However, quite surprisingly, the shift is not dramatic: $100$-anonymity
does not appear much more difficult to reach than $2$-anonymity.

\subsection{The bad: aggregation does not work}

\begin{figure}
\centering
\hspace*{-5pt}
\subfloat[d4d-civ]{
	\includegraphics[width=0.5\columnwidth]{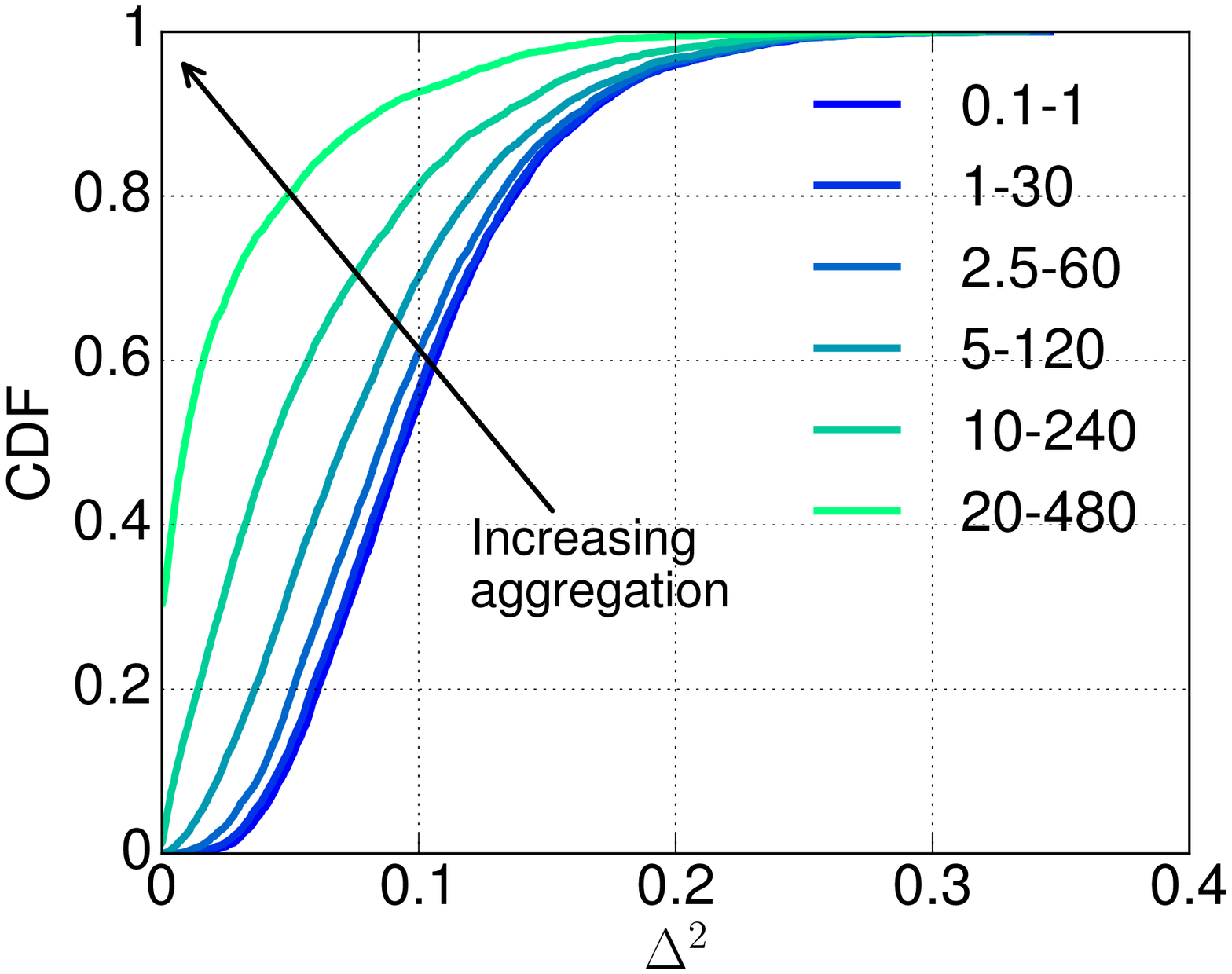}
	\label{fig:d4d_agg_civ}
}
\hspace*{-5pt}
\subfloat[d4d-sen]{
	\includegraphics[width=0.5\columnwidth]{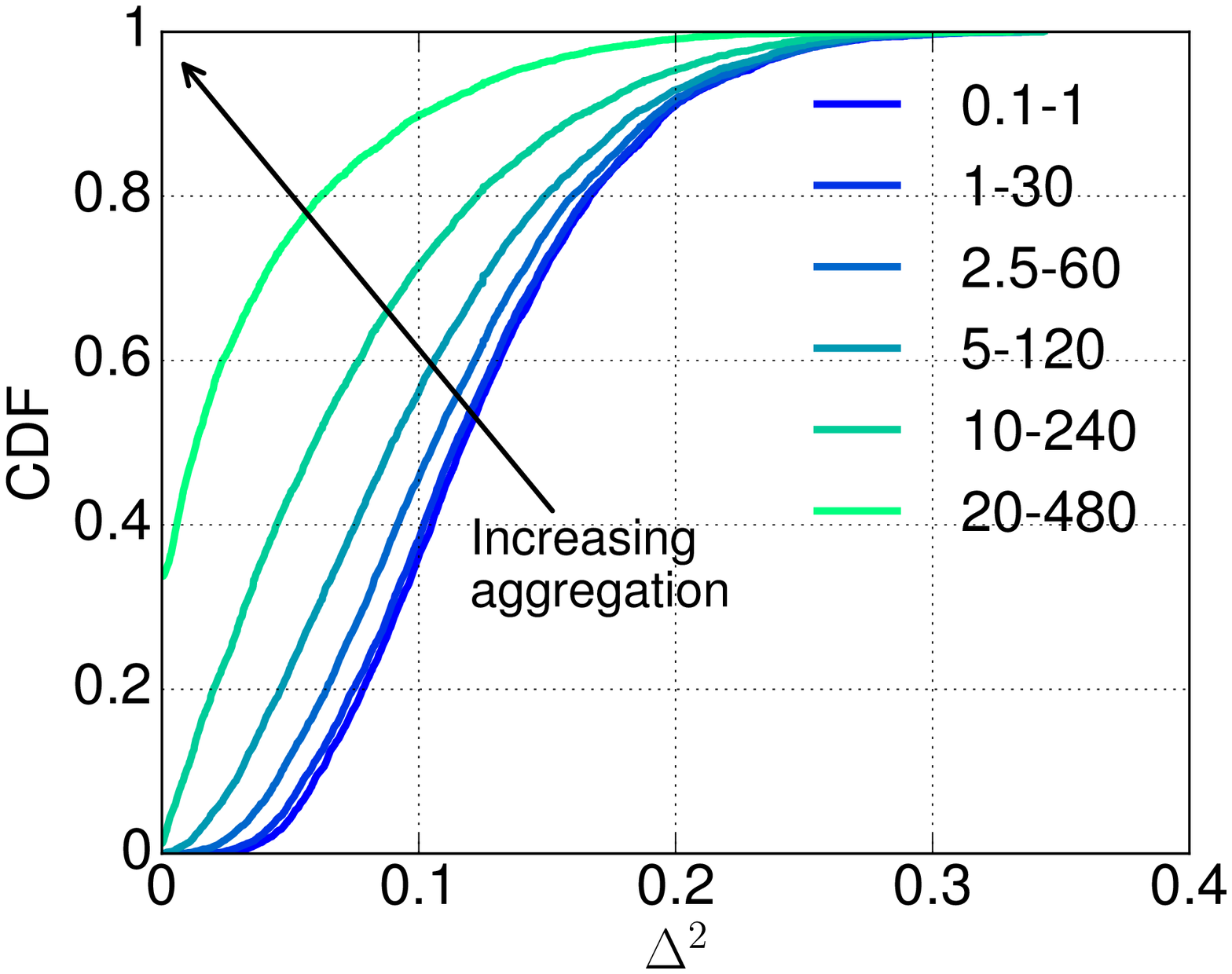}
	\label{fig:d4d_agg_sen}
}
\hspace*{-5pt}
\caption{CDF of the anonymizability measure, under the $2$-anonymity
criterion and for varying spatiotemporal aggregation levels, in the
\texttt{d4d-civ} and \texttt{d4d-sen} mobile traffic datasets.
The legend reports the level of spatial (in kilometers) and temporal
(in minutes) aggregation each curve refers to.
}
\label{fig:d4d_agg}
\end{figure}

Unfortunately, the easy anonymizability suggested by the distributions is
only apparent.
Fig.\,\ref{fig:d4d_agg} depicts the impact of spatiotemporal generalization
on anonymizability: each curve maps to a different level of aggregation, from
$100$~meters and $1$~minute (the finer granularity) to $20$~km and 8~hours.
As one could expect, the curves are pushed towards smaller values of the
anonymizability measure.
However, the reduction of spatiotemporal precision does not have the desired
magnitude, and even a coarse-grained citywide, 8-hour aggregation cannot
$2$-anonymize but 30\% of the mobile users.

This result is again in agreement with previous studies~\cite{zang11large,deMontjoye13unique},
and confirms that mobile traffic datasets are difficult to anonymize.

\begin{figure*}[tb]
\centering
\hspace*{-5pt}
\subfloat[d4d-civ, id $370$]{
	\includegraphics[width=0.16\textwidth]{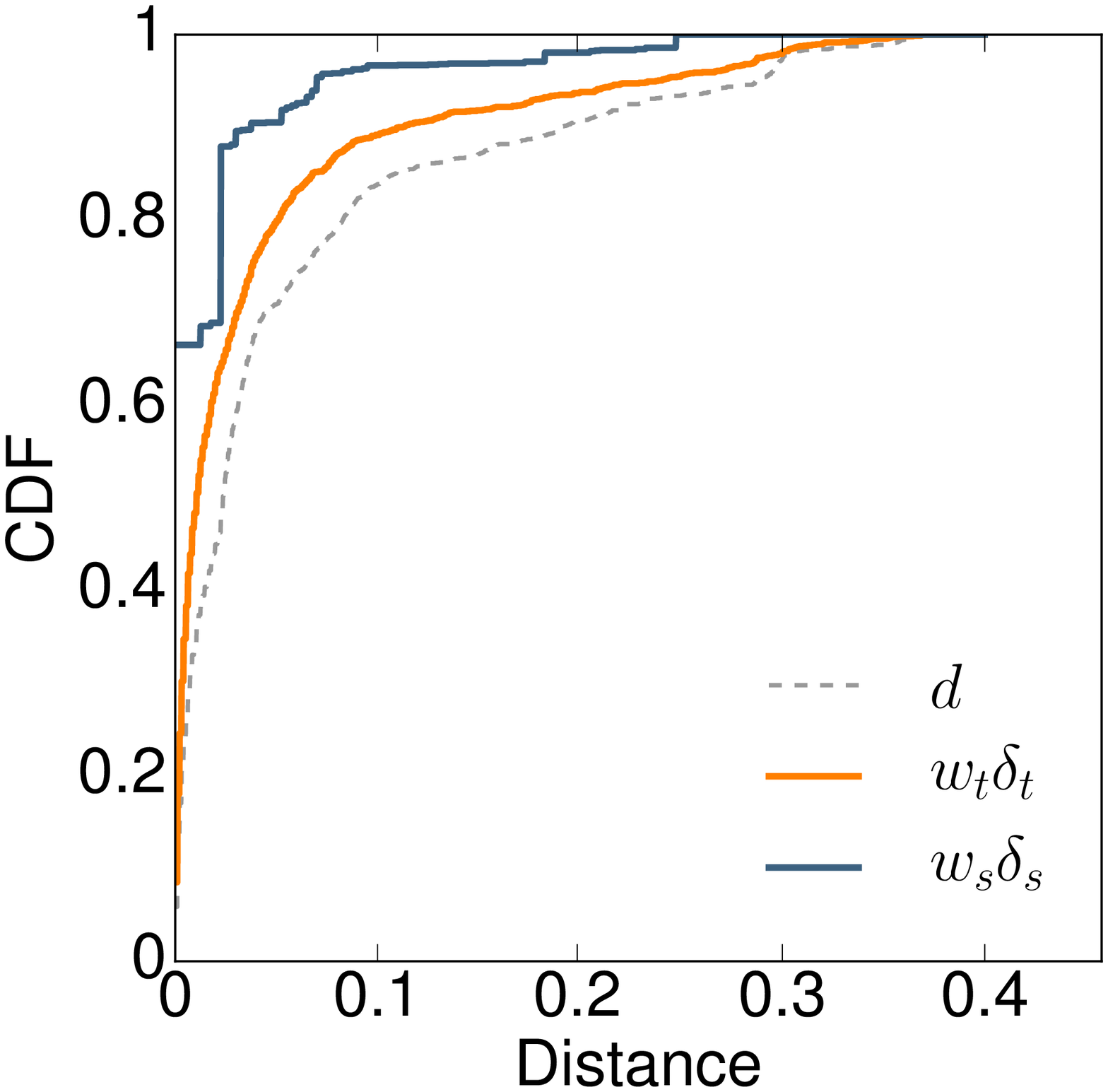}
	\label{fig:d4d_dis1}
}
\hspace*{-5pt}
\subfloat[d4d-civ, id $224$]{
	\includegraphics[width=0.16\textwidth]{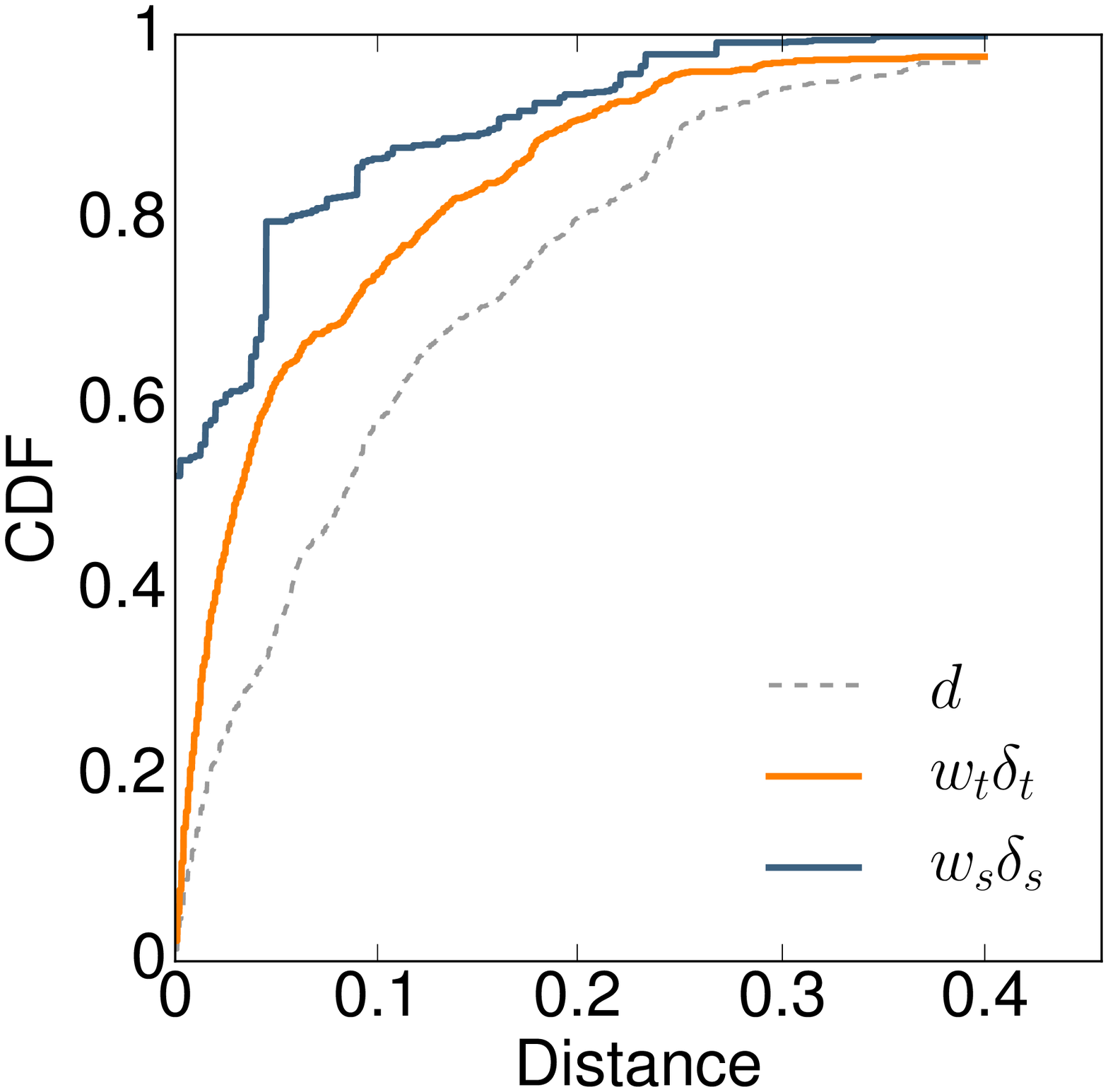}
	\label{fig:d4d_dis2}
}
\hspace*{-5pt}
\subfloat[d4d-civ, id $3175$]{
	\includegraphics[width=0.16\textwidth]{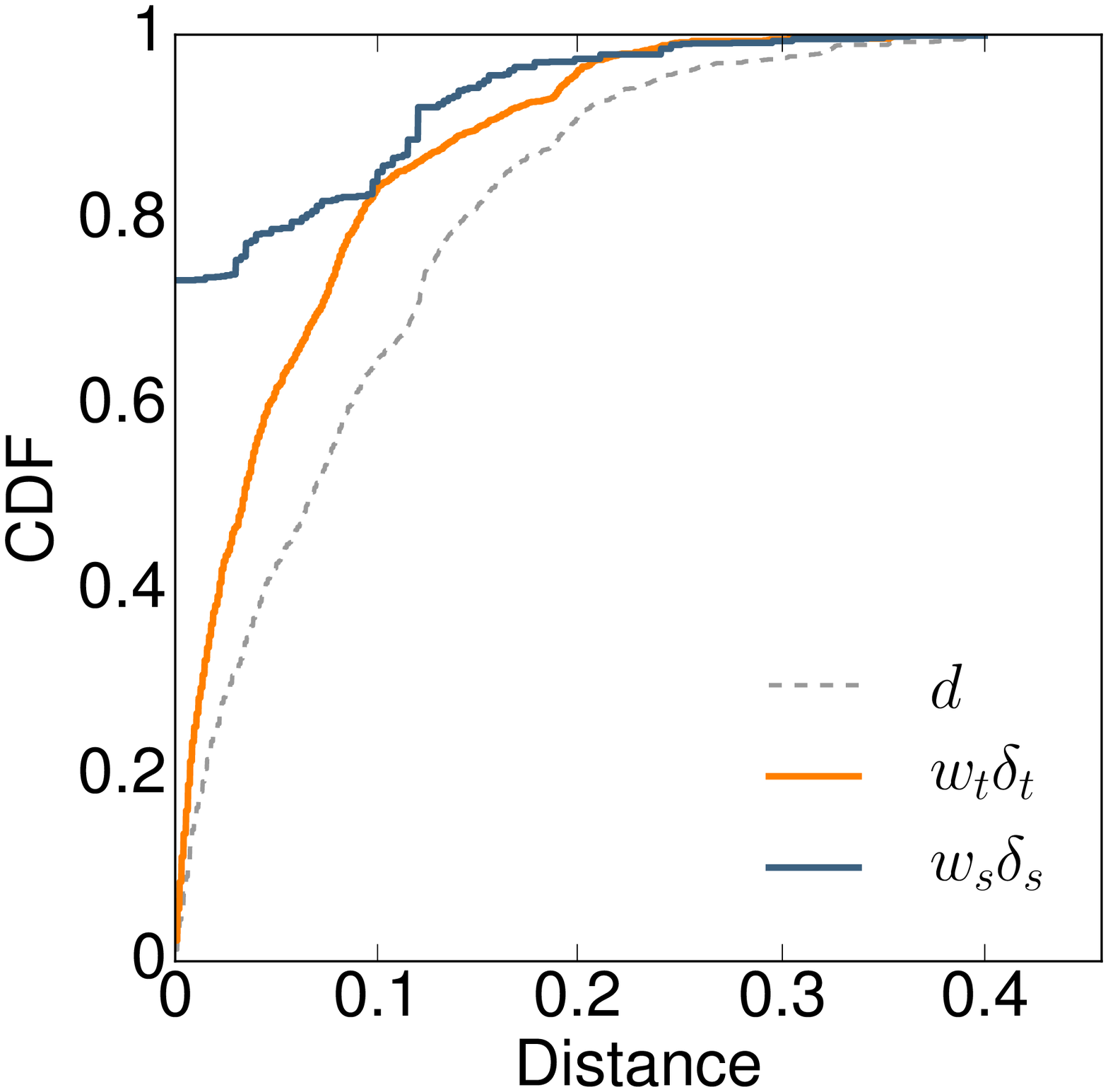}
	\label{fig:d4d_dis3}
}
\hspace*{-5pt}
\subfloat[d4d-sen, id $658$]{
	\includegraphics[width=0.16\textwidth]{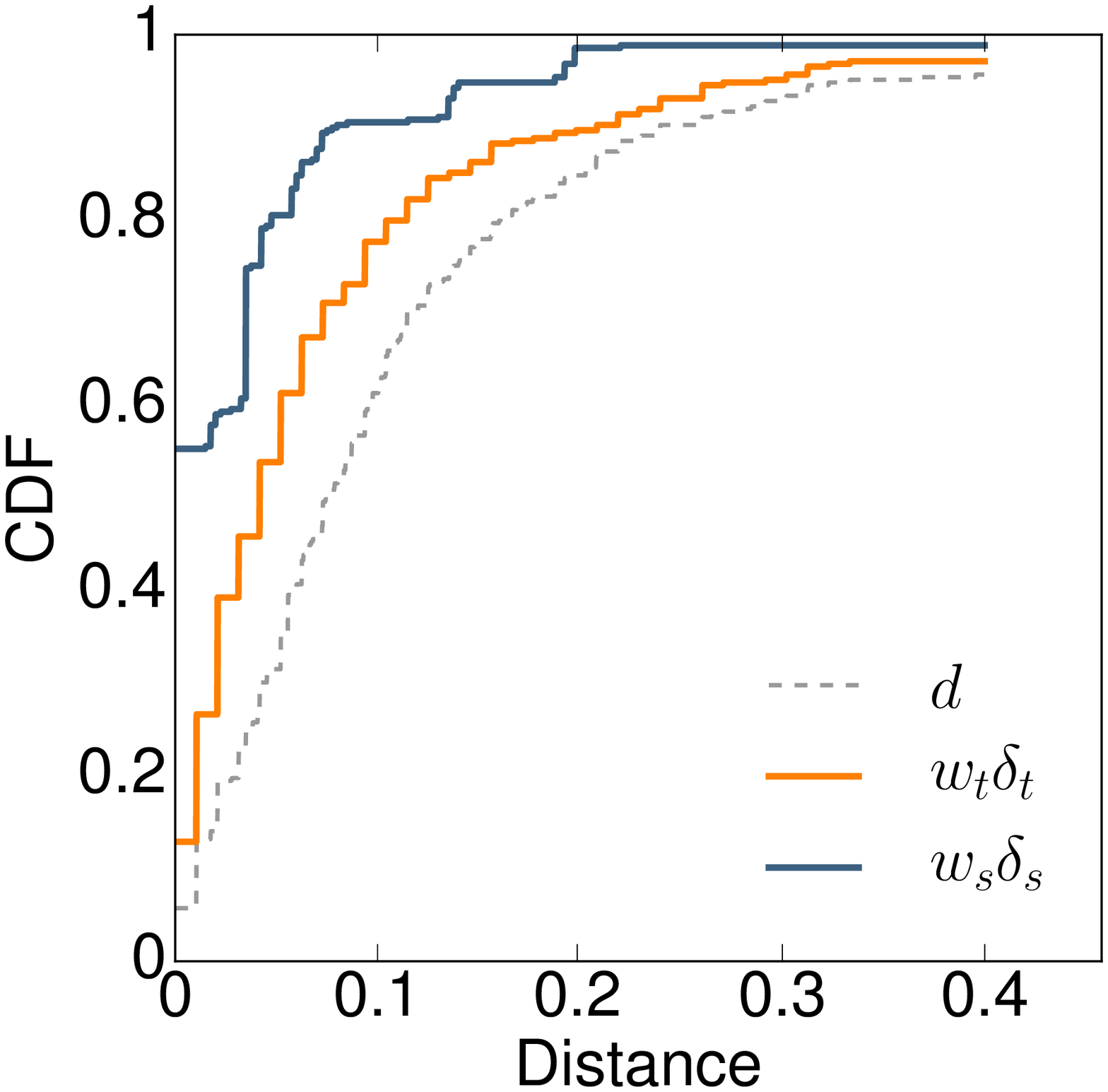}
	\label{fig:d4d_dis4}
}
\hspace*{-5pt}
\subfloat[d4d-sen, id $2130$]{
	\includegraphics[width=0.16\textwidth]{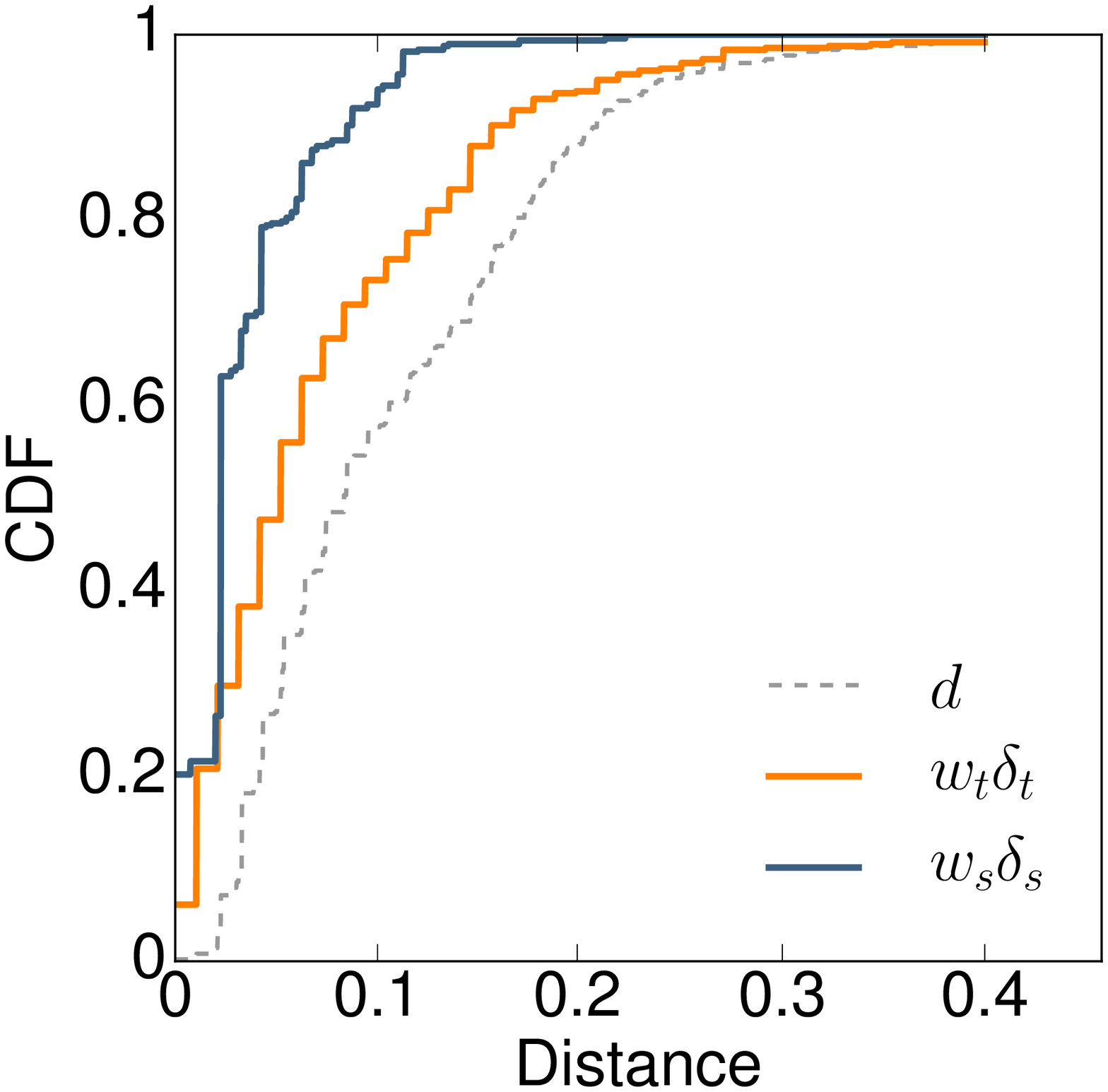}
	\label{fig:d4d_dis5}
}
\hspace*{-5pt}
\subfloat[Time weight]{
	\includegraphics[width=0.16\textwidth]{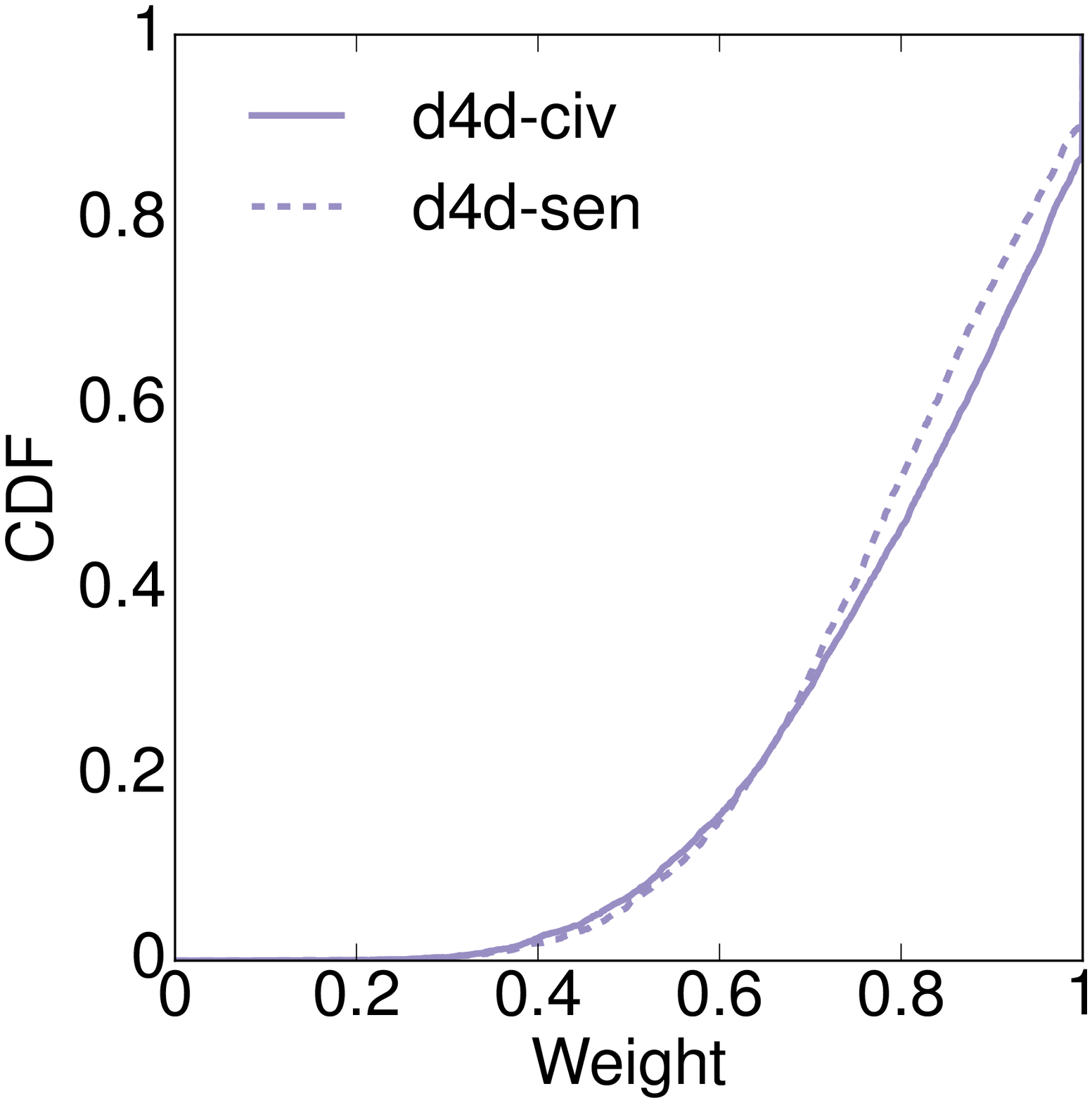}
	\label{fig:d4d_diswei}
}
\hspace*{-5pt}
\caption{(a)-(e) CDF of the sample distance, and of its spatial and
temporal components, under the $2$-anonymity criterion, for five
random mobile users in the \texttt{d4d-civ} and \texttt{d4d-sen}
mobile traffic datasets. (f) Contribution of the temporal components
to the total sample distance, expressed as the ratio between the sums
of temporal component distances and spatial component distances.}
\label{fig:d4d_ginitail}
\end{figure*}

\begin{figure*}[tb]
\centering
\hspace*{-5pt}
\subfloat[d4d-civ, Gini]{
	\includegraphics[width=0.24\textwidth]{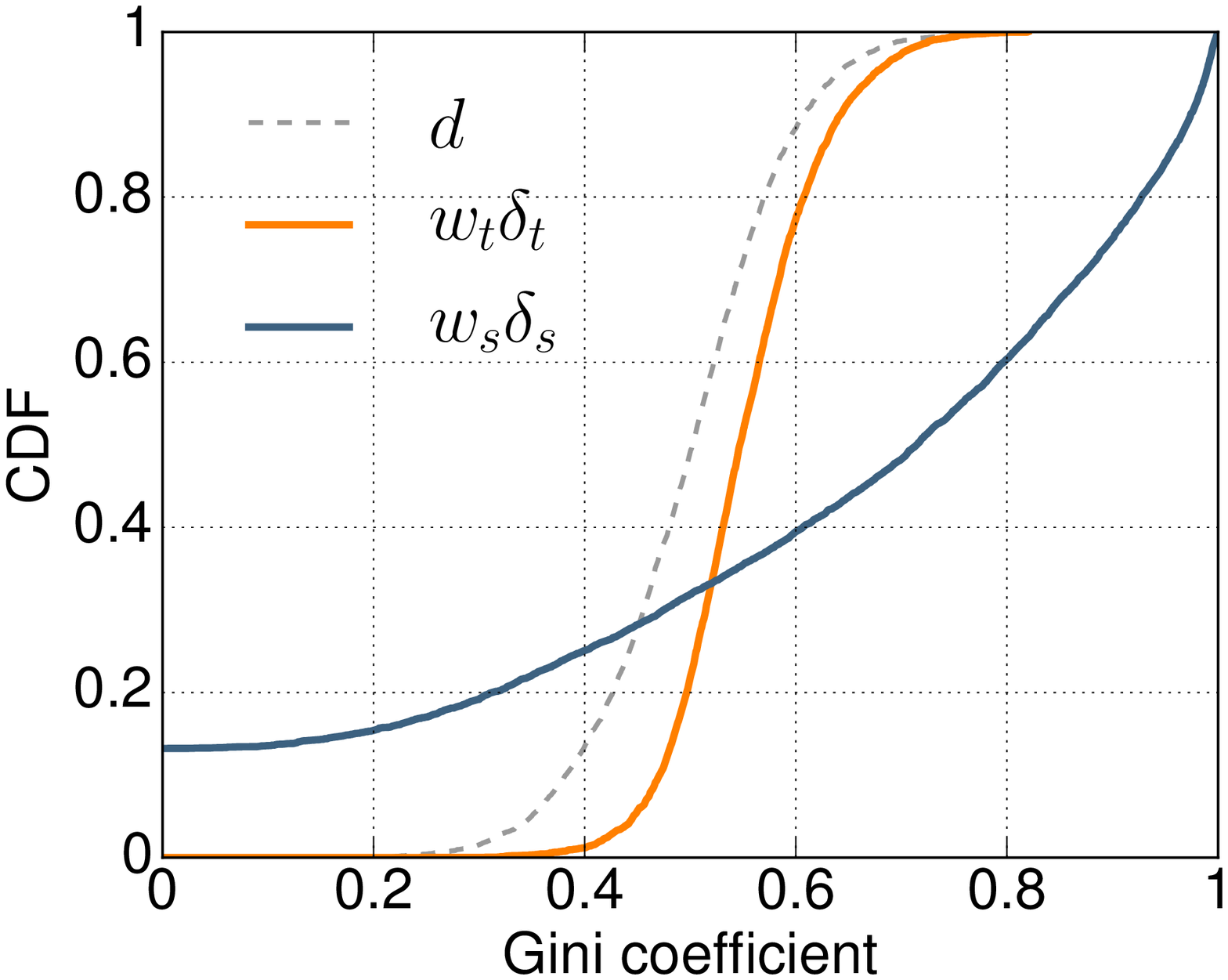}
	\label{fig:d4d_gini_civ}
}
\hspace*{-5pt}
\subfloat[d4d-civ, Tail weight]{
	\includegraphics[width=0.24\textwidth]{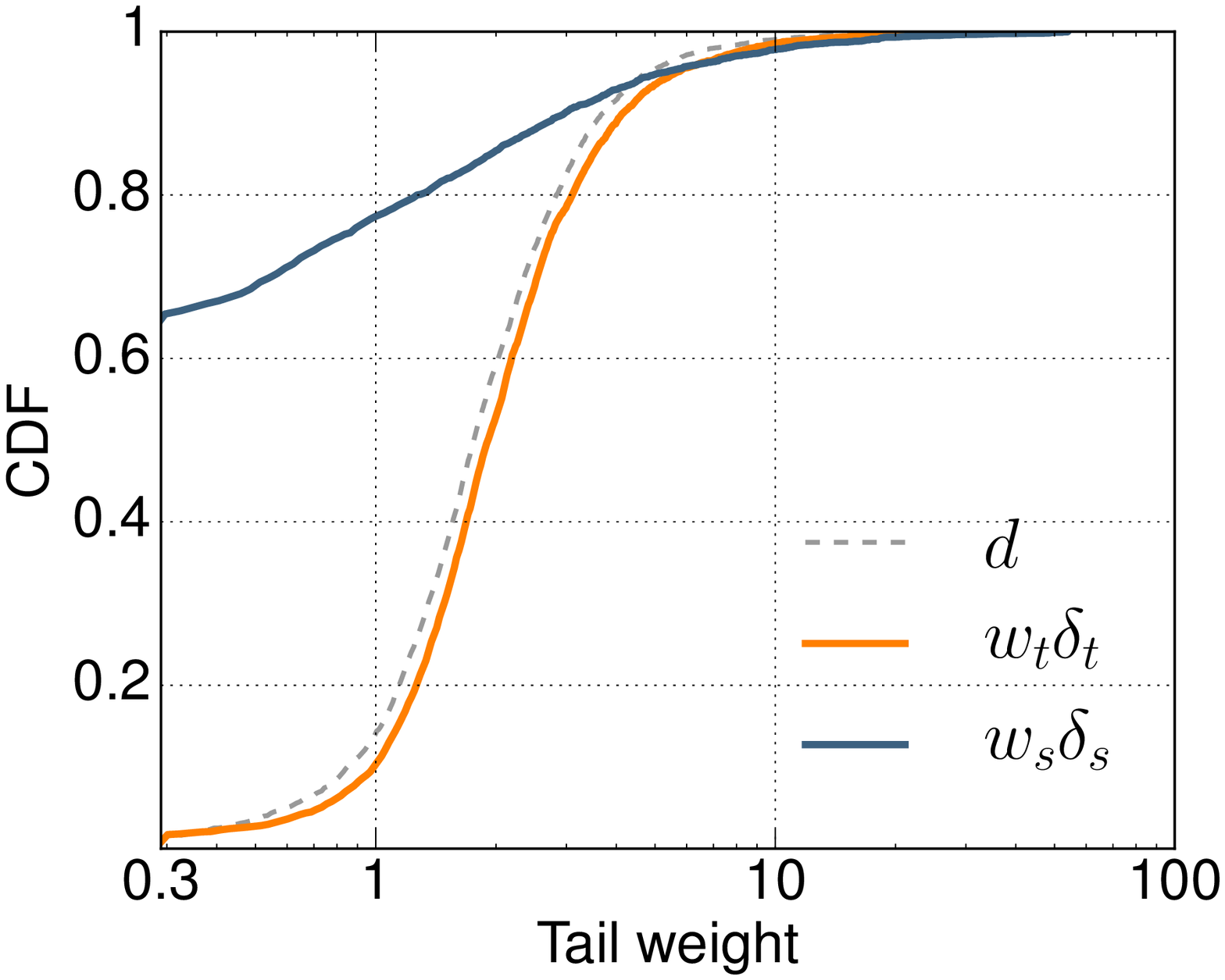}
	\label{fig:d4d_tail_civ}
}
\hspace*{-5pt}
\subfloat[d4d-sen, Gini]{
	\includegraphics[width=0.24\textwidth]{figures/d4d_civ_gini.eps}
	\label{fig:d4d_gini_sen}
}
\hspace*{-5pt}
\subfloat[d4d-sen, Tail weight]{
	\includegraphics[width=0.24\textwidth]{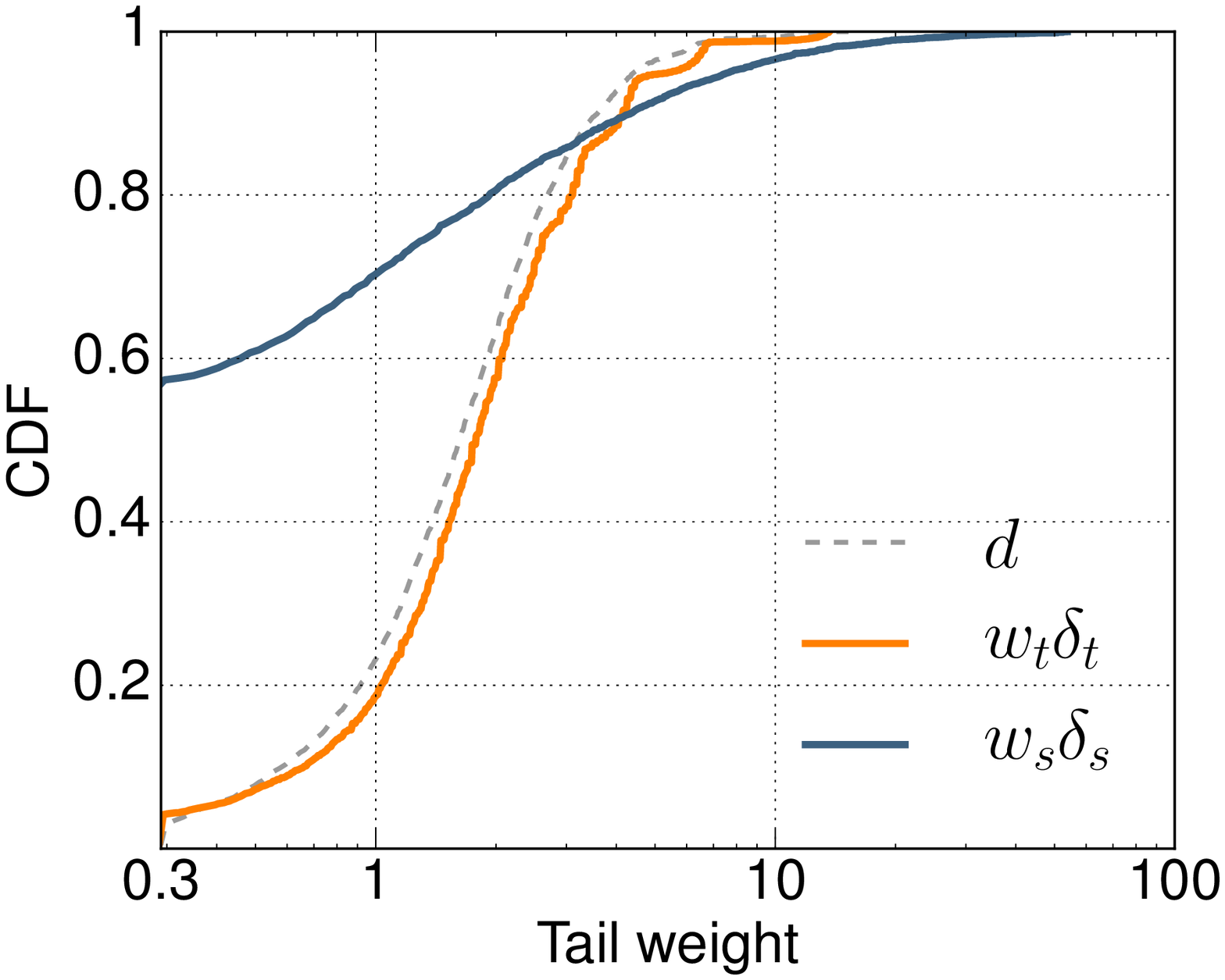}
	\label{fig:d4d_tail_sen}
}
\hspace*{-5pt}
\caption{(a,c) CDF of the Gini coefficient computed on the sample
distance distributions of all users in the \texttt{d4d-civ} and
\texttt{d4d-sen} datasets, for the $2$-anonymity criterion.
(b,d) CDF of the Tail weight index computed on the sample
distance distributions of all users in the \texttt{d4d-civ} and
\texttt{d4d-sen} datasets, for the $2$-anonymity criterion.}
\label{fig:d4d_ginitail}
\end{figure*}

\subsection{The why: long-tailed temporal diversity}

We are interested in understanding the reasons behind the incongruity above,
i.e., the fact that spatiotemporal aggregation yields such poor performance,
even if the average effort needed to attain $k$-anonymity is in theory not
elevate.

To attain our goal, we proceed along two directions. First, we separate
the spatial and temporal dimensions of the measure in (\ref{eqn:delta}),
so as to understand their precise contribution to the dataset anonymizability.
Second, we measure the statistical dispersion of the fingerprint distances
along the two dimensions: the rationale is that we observed the average
distance among fingerprints to be quite small, thus the reason of the low
anonymizability must lie in the deviation of sample distances around that mean.

\subsubsection{Impact of space and time dimensions}

Formally, we consider, for each user $a$ in the dataset, the set $\mathbb{N}^{k-1}_a$
of $k$-1 other subscribers whose fingerprints are the closest to that of $a$,
according to (\ref{eqn:delta}).
Then, we disaggregate all the fingerprint distances $\Delta_{ab}$ between $a$
and the users $b \in \mathbb{N}^{k-1}_a$ into sample distances $d_{ab}$, as per
(\ref{eqn:fingerdist}).
Finally, we separately collect the spatial and temporal components of all
such sample distances, in (\ref{eqn:sampledist}), into ordered sets
$\mathbb{S}^k_a = \{w_s\delta_s\}$ and $\mathbb{T}^k_a = \{w_t\delta_t\}$.
The resulting sets can be treated as disjoint distributions of the distances,
along the spatial and temporal dimensions, between the fingerprint of a generic
individual $a$ and those of the $k$-1 other users that show the most similar
patterns to his.

Examples of the spatial and temporal distance distributions we obtain in
the case of $2$-anonymity are shown in
Fig.\,\ref{fig:d4d_dis1}-\ref{fig:d4d_dis5}. Each plot refers
to one random user in the \texttt{d4d-civ} or \texttt{d4d-sen} dataset,
and portrays the CDF of the spatial ($w_s\delta_s$) and temporal ($w_t\delta_t$)
component distance, as well as that of the total sample distance ($d$).
We can remark that temporal components typically bring a significantly
larger contribution to the total fingerprint distance than spatial
ones. In fact, a significant portion of the spatial components is at
zero distance, i.e., is immediately $2$-anonymous in the original
dataset. The same is not true for the temporal components.

A rigorous confirmation is provided in Fig.\,\ref{fig:d4d_diswei},
which shows the distribution of the temporal-to-spatial component
ratios, i.e., $\sum_{\mathbb{T}^k_a} w_t\delta_t/\sum_{\mathbb{S}^k_a} w_s\delta_s$,
for all subscribers $a$ in the two reference datasets.
The CDF is skewed towards high values, and for half of mobile
subscribers in both \texttt{d4d-civ} or \texttt{d4d-sen} datasets
temporal components contribute to 80\% or more of the total sample
distance.
We conclude that the temporal component of a mobile traffic fingerprint
is much harder to anonymize than the spatial one. In other words,
{\it where} an individual generates mobile traffic activity is easily
masked, but hiding {\it when} he carries out such activity it is not so.

\subsubsection{Dispersion of fingerprint sample distances}

Not only temporal components weight much more than spatial ones
in the fingerprint distance, but they also seem to show longer
tails in Fig.\,\ref{fig:d4d_dis1}-\ref{fig:d4d_dis5}.
Longer tails imply the presence of more samples with a large
distance: this, in turn, significantly increases the level of
aggregation needed to achieve $k$-anyonimity, as the latter is
only granted once all samples in the fingerprint have zero distance
from those in the second fingerprint.

We rigorously evaluate the presence of a long tail of hard-to-anonymize
samples by means of two complementary metrics, still separating their
spatial and temporal components.
The first metric is the Gini coefficient, which measures the dispersion
of a distribution around its mean. Considering an ordered set
$\mathbb{S} = \{s_i\}$, the coefficient is computed as
\begin{equation}
G\left(\mathbb{S}\right) = 1 - \frac{2\sum_{i=1}^N i s_i + \sum_{i=1}^N s_i }{N\sum_{i=1}^N s_i},
\end{equation}
where $N$ is the cardinality of $\mathbb{S}$. We compute the Gini
coefficient on the sets $\mathbb{S}^k_a$ and $\mathbb{T}^k_a$, for
all users $a$.

The second metric is the Tail weight index~\cite{hoaglin1983understanding},
which quantifies the weight of the tail of a distribution with
empirical CDF $F$ as
\begin{equation}
\hspace*{-5pt}
T_{F} = \frac{F^{-1}\left(0.99\right)-F^{-1}\left(0.5\right)}{F^{-1}\left(0
.75\right)-F^{-1}\left(0.5\right)}
\frac{\Phi^{-1}\left(0.75\right)-\Phi^{-1}\left(0.5\right)}{\Phi^{-1}\left(0.9
9\right)-\Phi^{-1}\left(0.5\right)}.
\hspace*{-5pt}
\end{equation}
In the expression above, $F^{-1}(\cdot)$ is the inverse function of the empirical
CDF and $\Phi^{-1}(\cdot)$ is the inverse function of a standard normal CDF.
We compute again the Tail weight index on the distributions obtained from
both $\mathbb{S}^k_a$ and $\mathbb{T}^k_a$, for all $a$. 

Fig.\,\ref{fig:d4d_ginitail} shows the results returned by the two metrics
in the \texttt{d4d-civ} or \texttt{d4d-sen} datasets.
No significant differences emerge among the two mobile traffic datasets.
In both cases, the Gini coefficient, in Fig.\,\ref{fig:d4d_gini_sen} and
Fig.\,\ref{fig:d4d_gini_civ}, has, for all mobile user fingerprints ($d$),
high values around $0.5$ that denote significant dispersion around the mean.
However, two opposite behaviors are observed for the spatial ($w_s\delta_s$)
and temporal ($w_t\delta_t$) components.
The former show cases where no dispersion at all is recorded (coefficient
close to zero), and cases where the distribution is very sparse.
The latter has the same behavior as the overall distance, with values
clustered around $0.5$. The result (i) corroborates the observation that
the overall anonymizability is driven by distances along the temporal
dimension, and (ii) imputes the latter to the complete absence of
easy-to-anonymize short tails in the distribution of temporal distances.

Fig.\,\ref{fig:d4d_tail_sen} and Fig.\,\ref{fig:d4d_tail_civ} show instead
the CDF of Tail weight indices. Here, the result is even more clear: the
tail of temporal component distances is typically much longer than that of
spatial ones, and in between those of exponential and heavy-tailed distributions%
\footnote{As a reference, an exponential distribution with mean equal to 1
has a Tail weight index of 1.6, and a Pareto distribution with shape 1 has
an Tail weight index of 14.}.
Once more, the temporal component tail fundamentally shapes that of the overall
fingerprint distance.

\section{Discussion and conclusions}
\label{sec:conc}

At the light of all previous observations, we confirm the findings of
previous works on user privacy preservation in mobile traffic datasets.
Namely, the two datasets we analysed do not grant $k$-anonymity, not
even for the minimum $k$ = 2. Moreover, our reference datasets show
poor anonymizability, i.e., require important spatial and temporal
generalization in order to slightly improve user privacy.
The fact that these properties have been independently verified
across diverse datasets of mobile traffic suggests that
the elevate
uniqueness of trajectories
and low anonymizability are intrinsic
properties of this type of datasets.

In our case, even a citywide, 8-hour aggregation is not
sufficient to ensure complete $2$-anonymity to all subscribers.
The result is even worse than that observed in previous studies:
the difference is due to the fact that we consider the anonymization
of complete subscriber fingerprints, whereas past works focus on
simpler obfuscation of summaries~\cite{zang11large} or
subsets~\cite{deMontjoye13unique} of the fingerprints.

Our analysis also unveiled the reasons behind the poor anonymizability
of the mobile traffic datasets we consider, as follows.

On the one hand, the typical mobile user fingerprint in such datasets
is composed of many spatiotemporal samples that are easily hidden among
those of other users in the dataset.
This leads to fingerprints that appear easily anonymizable, since their
samples can be matched, {\it on average}, with minimal spatial and temporal
aggregation.

On the other hand, mobile traffic fingerprints tend to have a non-negligible
number of elements that are much more difficult to anonymize than the
average sample. These elements, which determine a characteristic dispersion
and long-tail behavior in the distribution of fingerprint sample distances,
are mainly due to a significant diversity along the temporal dimension.
In other words, mobile users may have similar spatial fingerprints, but
their temporal patterns typically contain a non-negligible number of
dissimilar points.

It is the presence of these hard-to-anonymize elements in the fingerprint
that makes spatiotemporal aggregation scarcely effective in attaining
anonymity. Indeed, in order to anonymize a user, one needs to aggregate
over space and time, until all his long-tail samples are hidden within
the fingerprints of other subscribers.
As a result, even significant reductions of granularity (and consequent
information losses) may not be sufficient to ensure
non-uniqueness
in mobile traffic datasets.

As a concluding remark, we recall that such uniqueness does not implies
direct identifiability of mobile users, which is much harder to achieve
and requires, in any case, cross-correlation with non-anonymized datasets.
Instead, uniqueness is a first step towards re-identification.
Understanding its nature can help developing mobile traffic datasets
that are even more privacy-preserving, and thus more easily accessible.

\end{document}